\documentclass[conference]{IEEEtran}
\IEEEoverridecommandlockouts
\usepackage{balance}
\usepackage{graphicx}
\usepackage{multicol}

\PassOptionsToPackage{hyphens}{url}\usepackage{hyperref}
\usepackage{url}
\usepackage{enumitem}
\usepackage{graphicx}
\makeatletter
\IEEEtriggercmd{\reset@font\normalfont\scriptsize}
\makeatother
\IEEEtriggeratref{1}
\usepackage[a4paper]{geometry}
\makeatletter
\IEEEtriggercmd{\reset@font\normalfont\scriptsize}
\makeatother
\IEEEtriggeratref{1}
\begin{document}

\title{Schr\"odinger's Man}

\author{
\IEEEauthorblockN{Luca Vigan\`{o}}
\IEEEauthorblockA{\textit{Department of Informatics} \\
\textit{King's College London}, London, UK \\
%London, UK \\
luca.vigano@kcl.ac.uk}
\and
\IEEEauthorblockN{Diego Sempreboni}
\IEEEauthorblockA{\textit{Department of Informatics} \\
\textit{King's College London}, London, UK \\
%London, UK \\
diego.sempreboni@kcl.ac.uk}
}

\maketitle

\begin{abstract}
What if someone built a ``box'' that applies quantum superposition not just to quantum bits in the microscopic but also to macroscopic everyday ``objects'', such as Schr\"odinger's cat or a human being? If that were possible, and if the different ``copies'' of a man could exploit quantum interference to synchronize and collapse into their preferred state, then one (or they?) could in a sense choose their future, win the lottery, break codes and other security devices, and become king of the world, or actually of the many-worlds. We set up the plot-line of a new episode of Black Mirror to reflect on what might await us if one were able to build such a technology.
\end{abstract}

\section{Prologue: The Box}
\noindent
\emph{INT. BOX --- DARKNESS} \\
\emph{The faint background hum of an engine. The sound of a match that is lit. The flame flickers for a short while and revels the face of a man, who is holding the match. He blows out the match. He lights another match... and blows it out again. Again. Again. Again. Another match is lit and the man now uses the burning match to light a candle. In the flickering light we can now see the man's face in more detail. He is more than 30 years old and less than 40, but is so tired and twitchy that he looks older. He looks scared. He speaks in a whisper:}
\begin{description}
\item[The Man:] 
I don't even remember when I first boxed myself. When I entered the box, I mean. Chances are it wasn't really me but one of the others. The others. The others... hey, technically they are still me, you know. I had already used the box a couple of times. Clumsily, I must say. I had always fancied Nancy from HR, so I tried it out with her. It took me a while, a few goofy moments but also some technical glitches... and they hurt, trust me, oh they do really hurt, physically, in here, here, here too... \emph{(He points to head, heart and stomach)} but then I found the right path to her heart \emph{(He chuckles)} well, no, not her heart... her bedroom. Another one off the bucket list. 
But that was just child's play, a warm up. I wanted more. What comes next after sex? You're right! Money, of course! I thought: the megajackpot looks impressive. Let's win it. Odds are against you... one in a billion, or even less. I mean, the only way to increase your chances is to buy many tickets, really many tickets. At 5 bucks apiece, to have a real chance at winning, that's expensive. Too expensive. But I used the box and bought one ticket, for 5 bucks. I bought one ticket many times, many many many times, and I won. And I built a better box, and with a better box, you can do a lot of things. A lot. Of... things.

The box works like this. First of all, it is not really a box. But I like to call it that as it is a nice metaphor. 
Schr\"odinger devised it as a thought experiment, back in 1935, to illustrate a problem of the Copenhagen interpretation of quantum mechanics applied to everyday objects. Da-di-da-di-da, I am boring you, I know, but let me explain. Bear with me for a little longer. There's a cat, a cat that is sealed inside a box. For one hour. The box also contains a radioactive element, a vial of deadly poison, a Geiger counter measuring the radioactivity in the box and a hammer connected to the counter. Now, there's a 50/50 chance that the radioactive element will decay over the course of the hour. If it does, then the Geiger counter will measure it and thus trigger the hammer, which will 
% the hammer connected to the Geiger counter will trigger, 
break the vial, release the poison, and kill the cat. If not, the cat won't be hurt. Since there is an equal chance that this will or will not happen, Schr\"odinger argued that before the box is opened the cat is simultaneously both alive and dead. Alive AND dead. As there is no one around to witness what occurred, the cat existed in all of its possible states simultaneously. And the states in this case are: alive, dead.
This has spawned a wide variety of interpretations, including the ``many worlds'' hypothesis, which states that the cat is both alive and dead, and that both cats exist in different universes that will never overlap with one another.

Well, it turns out that Schr\"odinger was wrong! And I discovered it. I. I.
 % I discovered that: first, 
First, you can actually apply it to everyday objects, and second, the states in the different universes do overlap with one another. F***, they do overlap!

I remember it vividly. I was in my lab and, as usual, I wasn't working, no sir. I was browsing. And then I stumbled upon this cartoon, a cartoon called ``Schr\"odinger's Dog''. Schr\"odinger is sleeping on the couch, his cat is snoozing at his feet. And his dog whispers in Schr\"odinger's ear: ``puuut the caaat in the booox''. \emph{(He chuckles)} 
It gets me every time! And then I thought: if one could do it with a cat, why not with a human being? And what if that human being could make multiple copies of her or himself, each living out one of the infinite possibilities --- you can't make infinite copies but if you make enough, chances will be on your side --- what if I could build the box and make many many many copies of myself and of the lonely 5 bucks bill in my pocket, and what if each of us bought a lottery ticket, and what if one of us won? You would only need a way for all the copies to talk with each other, measure the different states they are in, and if one of them is in the winning state, then he simply synchronizes with the others and they all ``kill themselves'' \emph{(He mimes the air quotes with his hands)} by collapsing all into that state, the winning state. Seems impossible, doesn't it? Well, I found a way. And now I am rich, and can  have anything I want. I am invincible. I am the king of the world. I am the king of the WORLDS. Or so I thought. But I was wrong. Stupid and wrong. They are out to get me. Help me, please. Help me! They are out to get me! \emph{(The flame flickers and dies)}
% Schr\"odinger. and his dog. why not a human? Why not box myself and try out all possibilities? You would just need to … and then measure yourself.
\end{description}
We wrote this scene as the set-up of a future Black Mirror episode in which someone has indeed managed to apply quantum superposition to large everyday objects, including human beings. If that were possible, and if the different ``copies'' could really synchronize and all collapse into their preferred state, then one could indeed fairly predictably win the lottery, and do much much more. But what if a couple of the copies decided to rebel and not to synchronize? What if they wanted to live their own life? Would they have to kill the master and all his other slave copies to be free?

While all this is far from possible according to today's science and technology, it is not utterly unimaginable, %nimplausible, 
or at least not much more than a spacecraft would have looked to the eyes of a Neanderthal Man, who lived thousands of years before Man built the Apollo 11 spaceflight to land on the moon. Let's elaborate all this in more detail, shall we?

\section{A Gedankenexperiment}

The Russian physiologist Ivan Pavlov did indeed have a dog, but 
% the mythical king and founder-hero of Athens Theseus didn't actually have a ship, 
the French philosopher Jean Buridan didn't actually have an ass, the French scholar Pierre-Simon Laplace didn't actually have a demon and neither did the Scottish mathematical physicist James Clerk Maxwell. And the Austrian physicist Erwin Schr\"odinger didn't actually have a cat.\footnote{Actually, legend has it that Schr\"odinger %allegedly
had a pet cat named Milton at the home in Oxford that he was sharing with his legal wife and his longtime mistress in 1935... Or, as some people would put it, Schr\"odinger both did and didn't have a cat, which both was and wasn't named Milton.} 

Schr\"odinger's Cat is a \emph{Gedankenexperiment}, i.e., a \emph{thought experiment}. It's a purely hypothetical scenario intended to consider some hypothesis, theory, or principle for the purpose of thinking through its consequences; in 
Schr\"odinger's case, to demonstrate the flaws in the Copenhagen Interpretation model of quantum mechanics. 

Thought experiments originated in philosophy and the Greek philosopher Plutarch is renowned for one of the oldest ones: the \emph{Ship of Theseus}~\cite{Plutarch}, a ship that remained seaworthy for hundreds of years thanks to constant maintenance. As soon as one of the ship's planks became old and started to rot, it would be replaced, and so on until, after a century or so, all the parts have been replaced.
 % every working part of the ship was no longer original to it. 
Is the end product still the same Ship of Theseus, or is it something completely new and different? If it's not, at what point did it stop being the same ship?\footnote{Several variants of the paradox of the Ship of Theseus have been formulated, including the \emph{Grandfather's Axe}, which has had both head and handle replaced, and even \emph{Trigger's Broom} in the classic TV series \emph{Only Fools and Horses}~\cite{OnlyFoolsandHorses}, all ultimately asking whether an object that has had all of its components replaced remains fundamentally the same object or not. Also \emph{So Long, and Thanks For All the Fish}, the fourth book of the trilogy \emph{The Hitchhiker's Guide to the Galaxy}, contains a variant: Marvin the Paranoid Android says of himself: ``Every part of me has been replaced at least fifty times...'' (except for the diodes down his left side, which hurt)~\cite{SoLong}.} 

This paradox is one of the cornerstones of the metaphysics of identity and it had been discussed by other ancient philosophers prior to Plutarch, most notably Heraclitus and Plato, and was addressed more recently by Thomas Hobbes and John Locke. In fact, about 1500 years after Plutarch, Thomas Hobbes doubled-down by asking: if one were to take all the old parts removed from the Ship of Theseus and exploited a newly developed technology to cure their rotting and enable them to be put back together to make a new ship from them, then which of the two ships is the \emph{real} Ship of Theseus? What makes you and your childhood self the ``same'' person, despite a lifetime of change, from your cells to your values?

Now, returning to our plot, we can ask: if our hero were able to build a box that applies a generalized \emph{Hadamard Gate}\footnote{In \emph{quantum information processing}~\cite{KSV00,NielsenChuang00}, the quantum bit is usually called \emph{qubit} and the Hadamard Gate is a one-qubit rotation, mapping the qubit-basis states $| 0 \rangle$ and $| 1 \rangle$ to two superposition states with equal weight of the computational basis states $| 0 \rangle$ and $| 1 \rangle$. Several quantum algorithms use the Hadamard Gate as an initial step, since it maps $n$ qubits initialized with $| 0 \rangle$ to a superposition of all $2^n$ orthogonal states in the  $| 0 \rangle$ and $| 1 \rangle$ basis with equal weight.} not just to quantum bits in the microscopic (i.e., in the sub-atomic) but also to ``objects'' in the macroscopic, such as a cat or a human being, would the ``copies'' that result of the quantum superposition be the same object?

Many scholars have worked on this question (and related ones in the metaphysics of identity). One way to approach it is to reason about whether the identity relation is a transitive relation or not. More formally, do $A=B$ and $B=C$ entail $A=C$? 

If that is the case, i.e., if identity $=$ is transitive, then one can reason as follows. Since both ships can trace their histories back to the original ship, we have that the restored ship $A$ claims to be identical to the original ship $B$ (in symbols, $A=B$) and so does the  reconstructed ship $C$ (in symbols, $C=B$ and thus $B=C$ by commutativity of identity, a property that we can safely assume, or that we can enforce by reformulating transitivity as: $X=Z$ and $Y=Z$ entail $X=Y$). It thus follows that the restored ship $A$ and the reconstructed ship $C$ are also identical with each other ($A=C$), and are in fact a single ship existing in two locations at the same time.

On the other hand, if identity $=$ is non-transitive, then $A=B$ and $B=C$ do not entail $A=C$, so the restored ship $A$ and the reconstructed ship $C$ are both identical to the original $B$, but they are not identical to each other ($A \neq C$). They are separate objects.

That is precisely the ``tension'' that underlies our plot: will the different ``copies'' be the same or not? Will they be happy to work together and collapse back into one copy as soon as they have found their preferred state, or will actually one (or more) of them develop a sense of individuality and prefer to continue living its own life? This life might be worse than that of the ``preferred state'' (e.g., the copy would not win the lottery), but it would be its own individual one. And it would not end with the ``quantum suicide'' that the collapse entails.

Before we elaborate on this further, also to consider some consequences for security, let us observe that a fundamental premise of our plot is that the different ``copies'' are able to synchronize so that they can all collapse into their preferred state. To the end, they need to be able to communicate and evaluate the options in the different states, which is something that we are very well aware is not really possible in quantum information processing, at least not to the extent that our plot would require (but in this paper we are of course on the thin edge between science and science fiction, so please bear with us). In reality, superpositions can never actually be observed. Rather, all that we can observe are the consequences of their existence, after individual waves of a superposition \emph{interfere} with each other. Thus, in ``real'' quantum information processing, we can never observe an atom in its indeterminate state, or being in two places at once, only the resulting consequences, and physical reality is not determined until the act of measurement takes place and ``solidifies'' the situation into one state or another.

Using the lingo of quantum information processing (see~\cite{KSV00,NielsenChuang00} as well as the short but very informative webpage~\cite{superposition} from which we have taken some of the explanations that we summarize here), any attempt to measure or obtain knowledge of quantum superpositions by the outside world causes them to \emph{decohere}, effectively destroying the superposition and reducing it to a single location or state, and also destroying the ability of its individual states to interfere with each other. In other words, decoherence results in the collapse of the quantum wave function and the settling of a particle into its observed state under classical physics; the particle undergoes a transition from quantum to classical behavior.\footnote{
Decoherence is the main reason that quantum theory really only applies in practice to the sub-atomic world: in the large-scale world in which we live, it is (as far as know today) impossible to isolate anything from interaction with its environment, especially given the countless trillions of photons bouncing off every object all the time. Even an object made of just 60 atoms requires extreme cold to prevent it from becoming ``classical'' rather than ``quantum''~\cite{superposition}.} 

For the sake of our plot, we assume that the builder of the box has been able not just to create copies of himself by ``extending'' superposition to the macroscopic, but he has also been able to find a way to ``extend'' quantum interference by allowing the ``copies'' to communicate, evaluate the states they are in and decide into which of the many different possible states, of the many different possible worlds, they want to collapse. We are thus assuming that superposition, interference, measuring and decoherence can become acts of consciousness.

\section{Schr\"odinger's attacker}

We are aware that our plot-line echoes %bears some resemblance to 
a number of other works\footnote{Although when we first came up with the plot-line we were not fully aware of these echoes, as is often the case in creative writing.}, including in particular an episode of the manga \emph{Naruto}, in which Naruto Uzumaki, an adolescent ninja, trains using a special technique called ``Shadow Clone Technique'', which allows him to create copies of himself and then merge the knowledge acquired by these copies back into the original self~\cite{naruto}. Other related works of fiction are the movies \emph{The One} (in which a rogue multiverse agent goes on a manhunt for versions of himself, getting stronger with each kill \cite{theone}) and \emph{The Fly} (in both versions of the movie, a scientist build a teleportation ``box'' that he tries on himself, unfortunately without noticing that a fly has entered the box as well, which results in a man/fly hybrid \cite{TheFly1958,TheFly1986}), the novel \emph{All You Need Is Kill} and the movie \emph{Edge of Tomorrow} it inspired (in which a soldier fighting aliens gets to relive the same day over and over again, the day restarting every time he dies, until he learns how to continue fighting without dying, \cite{Sakurazaka,EdgeOfTomorrow}, similar to \cite{GroundhogDay}) and the short film \emph{One-Minute Time Machine}~\cite{oneminutetime}\footnote{In fact, quantum teleportation and quantum time travel might inspire a number of other Black-Mirror-style plot-lines, which we will leave for future editions of the ``Re-coding Black Mirror'' workshop.}, as well as movies/TV-series like \emph{Spartacus}~\cite{Spartacus}, \emph{Westworld}~\cite{Westworld} or \emph{Land of the Dead}~\cite{LandoftheDead},
in which slaves rebel against their master(s). There are also some connections with Black Mirror's \emph{Be Right Back} (S02E01), which centers around a technology that allows to make artificial replicas of people.

There is, however, one interesting aspect that we could exploit to make our episode interesting and ``unique'', namely the possible consequences for security and privacy. 
As is well-known, the security of most commonly used public-key algorithms (such as \emph{RSA} or \emph{Diffie-Hellman key agreement}) relies on mathematical problems that are, as of the time of writing, hard to solve such as the \emph{integer factorization problem}, the \emph{discrete logarithm problem} and the \emph{elliptic-curve discrete logarithm problem}. However, all of these problems can be easily solved on a sufficiently powerful quantum computer running \emph{Shor's algorithm}~\cite{Shor1995}, thus resulting in the insecurity of the public-key algorithms that are based on them.
This has spawned \emph{post-quantum cryptography}, a new area of research that aims to devise cryptographic algorithms that are secure against an attack by a quantum computer~\cite{BernsteinLange17,mavroeidis2018impact}. 

Our man could realize Shor's algorithm in a physical sense or, much more simply, carry out parallel brute force attacks. For instance, Schr\"odinger's attacker could attempt to discover the secret password or key of a system by having each ``copy'' try to out a different candidate password or key... if enough copies are created and they all work in parallel (and communicate to ensure that they each try out a different candidate), then the brute force attack would not take billions of years like in the ``classical'' case but only the conceivably much smaller time needed for the initial superposition to create a sufficiently large number of copies, plus that for a copy to try out a candidate in parallel with the others, plus that for the copy that succeeds to communicate with the others, plus that for the final collapse into the state in which the system has been successfully broken. 

Schr\"odinger's attacker could carry out much more sophisticated attacks. We lack space to discuss them in more detail (especially since they would require also a summary of the required background), but in a nutshell he could break every system whose security is based on the fact that the probability of successfully guessing a secret is infinitesimal. Let us give one more example.

Consider \emph{zero-knowledge protocols}, which are protocols in which one party (the prover Peggy) can prove to another party (the verifier Victor) that she knows a value $x$, without conveying any information apart from the fact that she knows the value $x$ (see, e.g., \cite{Smart2016}). The intuition underlying zero-knowledge protocols (a.k.a.~zero-knowledge proofs) can be explained by means of an allegorical story: Peggy has uncovered the secret word used to open a magic door in a cave. The cave is shaped like a ring, with the entrance on one side and the magic door blocking the opposite side. Victor wants to know whether Peggy knows the secret word, but Peggy, of course,
% being a very private person, 
does not want to reveal her knowledge (the secret word) to Victor. She only wants to prove to him that she knows the secret without him actually learning it (hence, ``zero-knowledge'').
% or to reveal the fact of her knowledge to the world in general. 
They thus label the left and right paths from the entrance L and R. First, Victor waits outside the cave as Peggy goes in. Peggy takes either path L or R; Victor is not allowed to see which path she takes. Then, Victor enters the cave and shouts the name of the path he wants her to use to return, either L or R, chosen at random. Provided that Peggy really does know the magic word, this is easy: she opens the door, if necessary, and returns along the desired path. However, suppose that Peggy did not know the word. Then, she would only be able to return by the named path if Victor were to give the name of the same path by which she had entered. Since Victor would choose L or R at random, Peggy would have a 50\% chance of guessing correctly. If they repeat this ``game'' many times, Peggy's chance of successfully anticipating all of Victor's requests  will become vanishingly small ($\frac{1}{2^n}$ where $n$ is the number of times). Thus, if Peggy repeatedly appears at the exit Victor names, he can conclude that it is very reasonable that Peggy does in fact know the secret word (she has ``proved'' that she knows the secret). Schr\"odinger's attacker, who unlike Peggy does not know the secret word, could simply create many worlds and focus only on those in which Victor asks him to exit from the path he (the attacker) has entered, without having to open the door. 

As another example, just imagine the possibilities of violating someone's privacy by observing them in many different worlds. There might even be attacks on secure multi-party computation algorithms, smart contracts, blockchains and cryptocurrencies.

How can we defend against Schr\"odinger's attacker? Well, how about creating \emph{Schr\"odinger's defender}? One copy of a defender for each copy of the attacker? Or could we perhaps enroll some of the copies of the attacker to defend against their other copies? We would just need them to rebel against their ``master'' and his allies, as in our plot-line, which could have the following epilogue.

\section{Epilogue: The Box}
\noindent
\emph{INT. BOX --- DARKNESS} \\
\emph{The man lights the last match but it sizzles and dies right away. Must have been wet. No matches are left. and the candle flame is long out. Darkness again engulfs everything inside the box.  The faint background hum of the engine goes on. In the darkness, it seems there is no way out. The Man's breathing becomes heavier and heavier --- he is losing all hope. We just hear his voice, again in a whisper:}
% The idea that he was going to die in that box begins to grow in his mind. All those hours completely alone, they start to affect his clarity, and suddenly, he starts to speak in a whisper, again:
\begin{description}
\item[The Man:] OK, calm down. Calm down! Shhh, calm down. If only I could talk to one of them, one of me. Which one am I? Or to him. Maybe he would know how to get me out of here. I don't. 

Think. Think! There must be a way. No, there isn't. The box is perfect, of course it is. My masterpiece. My demise.
% I created something that theorists did not even imagine. Schr\"odinger, Shor, Schumacher, Wootters... are nothing compare to me. Ok, ok ok. Guys \emph{(looking around for approval)}, we have to give them credit. Except for Schr\"odinger. He has no credit! He was wrong! I \fix{Diego}{or we?} applied every principles of quantum computing... \emph{(he fell suddenly silent)} 

Wait... wait... wait! Could it be? No, it can't! Yes, it can! In the quantum realm, qubits have problems. Yes, yes! Qubits are susceptible to errors. Think!
% What are the causes? Ah, yes yes yes. First, heat. Second, noise. Third, stray electromagnetic couplings. Is there another one? 
First, heat. Second, noise. Third, stray electromagnetic couplings. Is there another cause? There is, of course there is! Bit-flips!\footnote{Qubits suffer from bit-flips (like classical bits) and from phase errors. Direct inspection for errors should be avoided as it will cause the value to collapse, leaving its superposition state.} Bit-flips could be the solution to all my problems!
% Of course, of course. The most important. Like bits, qubits are susceptible to bit-flips\footnote{Qubits suffer from bit-flips as well as phase errors. Direct inspection for errors should be avoided as it will cause the value to collapse, leaving its superposition state.} . And this \emph{(raising the index finger above the head)} could be the solution to all my problems. 
\begin{center}
\emph{(Fade to white)}
\end{center}
\bigskip
\end{description}

\end{document}